\documentclass[journal]{IEEEtran}
\ifCLASSINFOpdf
  \usepackage[pdftex]{graphicx}
\else
\fi
\usepackage[cmex10]{amsmath}
\begin{document}
\title{Large-angle scattering of multi-GeV muons on thin Lead targets} 
\author{A.~Longhin,~
        A.~Paoloni,~
	F.~Pupilli,~
        \\
        INFN Laboratori Nazionali di Frascati
\thanks{
A. Longhin, A. Paoloni, F.~Pupilli, INFN Laboratori Nazionali di Frascati, via Fermi 40, 00044 Frascati (RM), Italy.
andrea.longhin@lnf.infn.it
}
}
\markboth{}%
{A. Longhin \MakeLowercase{\textit{et al.}}: Large-angle scattering of muons on thin Lead targets}
\maketitle
\begin{abstract}
The probability of large-angle scattering for multi-GeV muons in thin
lead targets ($t/X_0$ of ${\mathcal{O}}(10^{-1})$) is studied. The new
estimates presented here are based both on simulation programs (GEANT4
libraries) and theoretical calculations. In order to validate the
results provided by simulation, a comparison is drawn with
experimental data from the literature.  This study is particularly
relevant when applied to muons originating from $\nu_\mu$ CC
interactions of CNGS beam neutrinos. In that circumstance the process
under study represents the main background for the
$\nu_\mu\to\nu_\tau$ search in the $\tau \to \mu$ channel for the
OPERA experiment at LNGS.  Finally we also investigate, in the CNGS
context, possible contributions from the muon photo-nuclear process
which might in principle also produce a large-angle muon scattering
signature in the detector.
\end{abstract}


\section{Introduction}
\IEEEPARstart{M}{ultiple} Coulomb scattering has been studied
extensively in the past in many different conditions and for different
incident particles both theoretically and experimentally. In this work
we are interested in the characterisation of large-angle Multiple
Coulomb Scattering (LAS) of multi-GeV muons scattering over high-$Z$
materials with a thickness being a fraction of the radiation length
$X_0$ (i.e. thickness ${\mathcal{O}}$(1 mm)). In the LAS regime we are
interested in, the transferred momentum $q$ is such that the De
Broglie wavelength ($\lambda \simeq \hbar/q$) of the probe is much
smaller ($< 1$ fm) than the typical radial extension of a high-$Z$
nucleus (about 6-7~fm for Lead).  For a correct description of the
high-angle tails of the scattering angle distribution it is thus
mandatory to precisely take into account the effect of the nuclear
charge distribution as well as the contribution of scattering off
individual protons.

In Sect.~\ref{sec:theory} we review basic theoretical aspects of the
problem highlighting the associated challenges.  In
Sect.~\ref{sec:MCsim} we overview the available approaches for
simulating multiple Coulomb scattering while in Sect.~\ref{sec:implem}
we describe specifically the simulation program developed for this
work using GEANT4 (\cite{GEANT4A}, \cite{GEANT4B}) libraries. In
Sect.~\ref{sec:G4theo} we compare the developed simulation with a
theoretical approach described in Sect.~\ref{sec:theory}.  In
Sect.~\ref{sec:G4data} we validate the simulation on experimental data
coming from past experiments designed to study the nuclear charge
density.  Finally, in Sect.~\ref{sec:OPERA}, the impact of LAS as a
background for the $\nu_\mu \to \nu_\tau$ search is evaluated.  The
importance of muon photo-nuclear interactions in mimicking LAS
topologies at CNGS is evaluated in Sect.~\ref{sec:photonuc}.

\section{Multiple scattering theories}
\label{sec:theory}

Multiple scattering has been an active field of theoretical and
experimental efforts since very early days of particle physics.
Goudmit-Saunderson~\cite{GS,GS1} and Lewis~\cite{Lewis} developed a
theory which allows to calculate the exact distribution of the angle
in space ($\theta$) due to multiple scattering after a given path
$s$. The problem was tackled by formulating it in terms of a transport
equation solved by expanding the angular dependence in Legendre
polinomials ($P_l(\cos\theta)$). The angular distribution is found to
be given by:
\begin{equation}
F(\theta,s)=\sum_{l=0}^{\infty}\frac{2l+1}{4\pi}e^{-{\frac{s}{\lambda_l}}}P_l(\cos\theta)
\end{equation}
with the transport mean free paths defined as:
\begin{equation}
{\lambda_l}^{-1}=e^{-2\pi s N\int_{-1}^{1}{(1-P_l(\cos x))\frac{d\sigma(x)}{d\Omega}}d(\cos x)}.
\end{equation}
The theory does not assume any particular form of the single
scattering differential cross section $d\sigma/d\Omega$ and provides
exact solutions. It can also deal with arbitrary nuclear form factors
provided that integrals are evaluated numerically.  Nevertheless for
large scattering angles the calculation requires the evaluation of
high order Legendre polinomials which are difficult to treat. This
limitation is also relevant for the present case and hence this
approach was not considered further.

Another interesting theoretical study was performed in~\cite{Meyer} by
Meyer. The theory by Moli\`ere~\cite{Moliere,Moliere1} is there
extended to take into account a generic nuclear form factor. The
general distribution of the scattering angle $\theta$ is derived and,
for a point-like nucleus, it reads as:
\begin{equation}
F(\theta,t)\theta d\theta = 2\chi e^{-\chi^2}\left[ 1 +\frac{b_0+b_2\chi^2+\sum_{\nu=2}^{\infty}{b_{2\nu}\chi^{2\nu}}}{B} \right]d\chi 
\label{eq:eq1}
\end{equation}
$\chi$ being equal to $\theta$ apart for numerical constants which
depend on the parameters of the problem (incoming momentum, particle,
target thickness, density etc.) which are made explicit in Appendix A.
The other quantities are either constants $b_0 = 0.423$, $b_2=-1.423$,
$b_{2\nu}=\frac{(\nu-2)!}{(\nu!)^2}$ or simple functions of the system
parameters as for $B$ (see Appendix A for more details).  The
component having the form $\chi e^{-\chi^2}$ results from multiple
consecutive scattering in the target thickness while the tail at
higher $\chi$ arises from the occurrence of single high angle
scattering.  For a finite nucleus with a charge density distribution
$q(x)$ the $b_{0,2,2\nu}$ coefficients in Eq.~\ref{eq:eq1} are
replaced by other coefficients $c_{0,2,2\nu}$ involving integrals of
the form factor $q(x)$ (see Appendix A).  If we approximate the charge
density of the nucleus as a uniform sphere the resulting form factor
takes this form:
\begin{equation}
q(x) = \frac{9}{x}\left[\frac{\sin{x}}{x}-\cos x \right]
\label{eq:eq3}
\end{equation}
As a further simplification we can approximate $q(x)$ with a
Gaussian function $q=e^{-a x^2}$.  The coefficients $c_{0,2,2\nu}$ can then be
expressed analitically as simple functions of $Z$, $a$ and the nuclear radius
(Appendix A). Unfortunately also this method faces numerical limitations
that we will discuss in Sect.~\ref{sec:G4theo}. It will anyway be used
for small scattering angles as a validation for the Monte Carlo
approach which we are about to describe in Sect.~\ref{sec:MCsim}.

\section{Monte Carlo simulation of multiple scattering}
\label{sec:MCsim}

Multiple Coulomb scattering can be simulated numerically using different approaches: 
\begin{itemize}
\item \emph{detailed simulations}: each track is simulated
  at microscopic level as a succession of connected straight segments (free-flights),
Changes of direction are determined by sampling the polar deflection
$\cos\theta$ from the single scattering distribution, and the
azimuthal scattering angle uniformly in [0, 2$\pi$]. The length $t$ of
the free-flight is sampled from the exponential distribution
$p(t)=\lambda^{-1}e^{-t/\lambda}$.  The simulated angular and spatial
displacements are ``exact'', i.e. they coincide with those obtained
from a rigorous solution of the problem (i.e. by mean of the transport
equation introduced in the theories of Goudsmit-Saudersons~\cite{GS,GS1}
and Lewis~\cite{Lewis}).
Nevertheless detailed simulations are in practice
not possible computationally except for specific cases involving very
low energies and thicknesses.

\item \emph{condensed simulations}: the global effects of the
  collisions during a macroscopic step are simulated using
  approximations. Usually the theories of
  Goudsmit-Saundersons~\cite{GS,GS1}, Moli\`ere~\cite{Moliere,Moliere1}
  and Lewis~\cite{Lewis} are used to predict the angular and spatial
  distributions after each step.  At energies where condensed
  algorithms are required, the great majority of elastic collisions
  are ``soft'' collisions with very small deflections. Especially the
  spatial distributions might present significant dependencies on the
  choice of the macroscopic step which has to be carefully studied.
\item \emph{mixed algorithms}: 
  Hard collisions, with scattering angle $\theta$ larger than a given
  value $\theta_S$, are individually simulated and soft collisions
  (with $\theta<\theta_S$) are described by means of a multiple
  scattering approach. It is clear that, by selecting a conveniently
  large value for the cutoff angle $\theta_S$, the number of hard
  collisions per electron track can be made small enough to allow
  their detailed simulation. The value of the mean free path between
  the simulation of hard collisions, $\lambda_{(h)}$, is given by
  $\lambda_{(h)}^{-1}=2\pi N  \int_{\theta_S}^{\pi}\sigma(\theta)\sin\theta d\theta$.
  As the fluctuations in the spatial displacement after a certain path
  length are mainly due to hard collisions, a mixed procedure (with a
  suitably small value of $\theta_S$) yields spatial and angular
  distributions that are more accurate than those from a condensed
  simulation. 
\end{itemize}

Since we are interested in rare high-$q$ events a mixed algorithm
approach has been preferred. Furthermore in this scheme the inclusion
of nuclear effects can be easily taken into account when sampling the
angular distribution at single-scattering level.

\section{The \textsc{Geant4} Monte Carlo implementation}
\label{sec:implem}
\begin{figure}[!t]
\centering
\includegraphics[width=9cm]{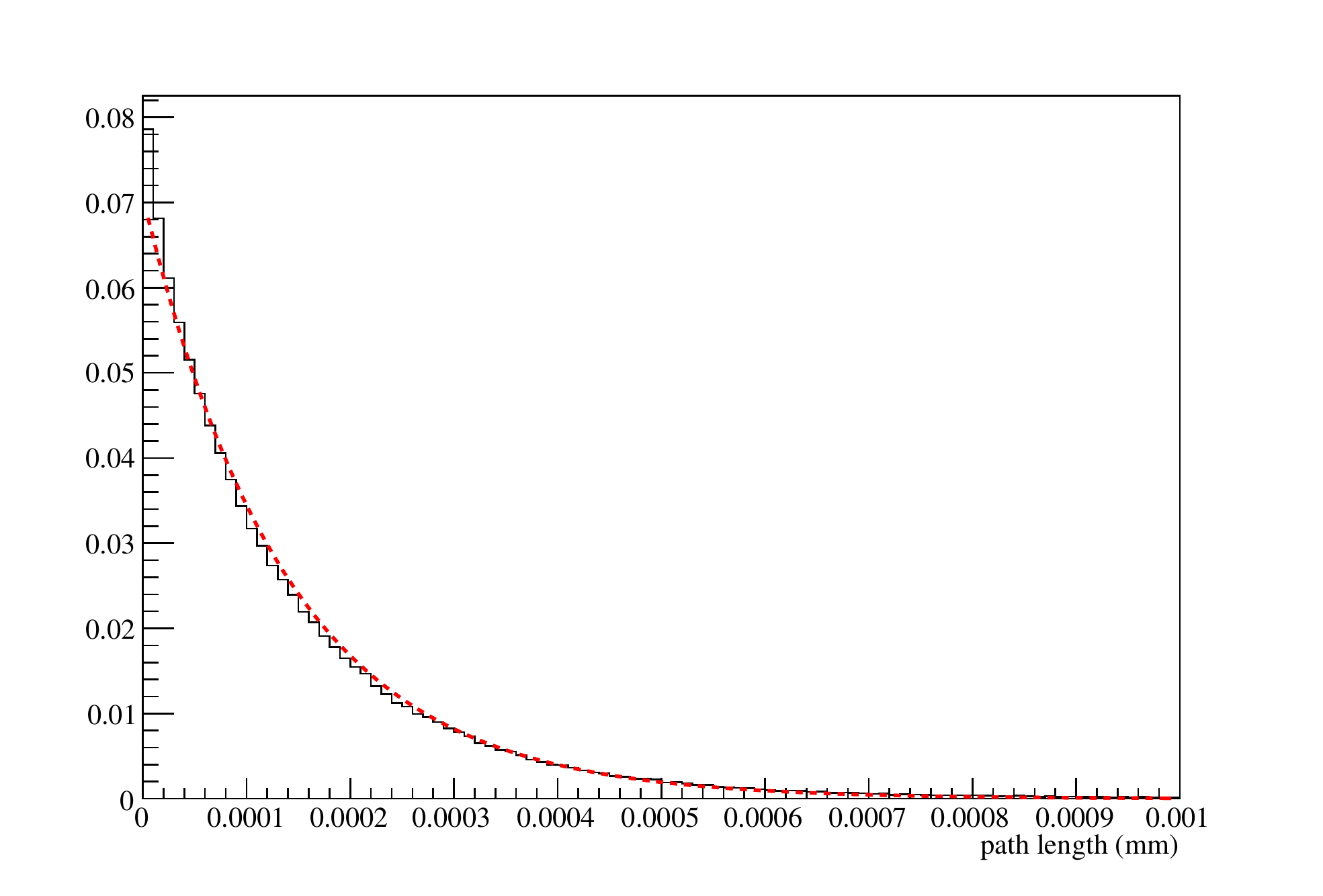}
\caption{Distribution of the step length for the GEANT4 mixed simulation for 2~GeV/c muons over 1 mm of Lead.}
\label{fig:steps}
\end{figure}

The version of \textsc{Geant4} used for this study is
{\textsc{4.9.6.p02}} with the \verb+standardSS+ physics list option to
enable the treatment of single scattering above a certain angular
threshold\footnote{``.mac'' file setting: /phys/addPhysics
  standardSS.}.  The simulation of single-scattering is dealt with by
the routine \verb&G4WentzelOKandVIxSection.cc&.
The distribution of the step length used by the GEANT4
simulation is shown in Fig.~\ref{fig:steps} for 2~GeV/c muons on 1~mm
thick lead target. The slope of the exponential distribution
corresponds to a value of $\lambda_{(h)} = 0.14~\mu$m such that the
single-scattering code is invoked on average about 7000 times for each
muon crossing the 1mm thick lead target.

The event sampling is done according to the model by Wentzel~\cite{ref:Wentzel} in the formulation
of~\cite{ref:Fernandez}. The model adopts a modified Coulomb potential
to take into account the effect of the screening of the electron
cloud: $V(r)=(Ze^2/r)e^{-r/R}$.  A slightly modified version of the
Moli\`ere screening angle $\chi_a^\prime=(\frac{\hbar}{2pR})^2$ is used whose
$Z$ and energy dependence is parametrised as
(cfr.~\cite{ref:Fernandez}, eq. 23):
\begin{equation}
\chi_a^\prime= \frac{\alpha^2 {a_0^2Z^{2/3}}}{p^2} \left(1.13+3.76 \left(\frac{\alpha Z}{\beta}\right)^2\sqrt{\frac{\tau}{\tau + Z^{2/3}}}\right)
\end{equation}
$a_0$ being the Bohr radius and $\tau=E_{kin}/m$.

If we define $\mu = \frac{1}{2}(1-\cos\theta)$ the differential cross
section $d\sigma/d\mu$ in presence of screening is proportional to
$\sim p^{-2}(\chi_a^\prime+\mu)^{-2}$. The inverse transform method leads to
the following sampling formula:
\begin{equation}
 \mu = \mu_S +\frac{(\chi_a^\prime+\mu_S)\zeta(1-\mu_S)}{(\chi_a^\prime+1)-\zeta(1-\mu_S)} \simeq \frac{\mu_S}{1-\zeta(1-\mu_S)}
\label{eq:sampling}
\end{equation}
with $\mu_S= \frac{1}{2}(1-\cos\theta_S)$. In the multi-GeV regime the
screening parameter $\chi_a^\prime$ is found to be of
${\mathcal{O}}(10^{-12})$ and hence the sampled distribution
effectively reduces to a Rutherford law $dN/d\theta \propto
(1-\cos\theta)^{-2}$.
The single scattering angular distribution generated from
Eq.\ref{eq:sampling} is then corrected for effects related to the
atomic electrons, the nuclear recoil and the nuclear charge density by
introducing a function $g_{rej}(\theta) = g_{a}(\theta)\times g_{rec}(\theta) \times \vert
F_N(\theta)\vert^2$.  A random number $\zeta^\prime$
is thrown between 0 and 1 and the scattering is discarded if
$\zeta^\prime>g_{rej}$.
The atomic correction reads
$g_a(\theta)=1-\beta^2\sin^2\frac{\theta}{2} + \frac{\alpha\beta \pi
  Z}{2} (1+\cos\theta)\sin{\frac{\theta}{2}}$ while the recoil
correction follows the term found in the Mott scattering formula
$g_{rec}=\left(1 + (1-\cos\theta) \frac{p}{M}\right)^{-1}$.

In the standard implementation nuclear charge effects are implemented through a dipole form factor 
\mbox{$\vert F_N(q)\vert^2 = {(1+\frac{1}{12}(\frac{qR_N}{197})^2)}^{-2}$} with $q\simeq p\theta$ in MeV. The nuclear
extension is parametrised as $R_N=1.27A^{0.27}$ giving $R_N=5.37$~fm for the $^{208}$Pb nucleus.
Dipole form factors arise from an exponentially decreasing nuclear charge density and are thus not
appropriate to describe the charge distribution of large nuclei.
We have then modified the simulation program by introducing an ad~hoc form factor for Lead based on the 
Saxon-Woods parametrisation of the charge density which is known to produce a much more realistic
description:
\begin{equation}
\rho_{SW}(r)= \rho_0\left(1+e^{\frac{r-b}{a}}\right)^{-1}
\label{eq:SW}
\end{equation}
The Saxon-Woods form factor has been fitted on precise electron
scattering data available in the literature~\cite{Frois}. We have used
$b = 6.647$ fm and \mbox{$a=2.30/(4\ln 3)$~fm} as
in~\cite{ref:parametriSW}.  The dipole form factor predicts a largely
overestimated contribution of high-$q$ scatterings due to the
assumption of a charge density being much more concentrated with
respect to what is actually observed. It is hence only appropriate for
small $q$.  The form factor $\vert F_N^{SW}(q)\vert^2$ has been then
extracted numerically by taking the squared Fourier transformation of
the charge density $\rho_{SW}(r)$. The analytical dipole formula for
$\vert F_N(q)\vert^2$ has been then replaced by a look-up table
representing the Saxon-Wood form-factor of in bins of $q$ (100 bins
from $q=0$ to $q=788$ MeV/c). Values above 788~MeV have been
neglected.
We have also considered the contribution of scattering on single
protons ($F_p$) by composing the two effects according to the relative
charge $Z$ following the procedure of Butkevic et al. (\cite{ref:Butkevich}):
\begin{equation}
\vert F(q)\vert^2 = \vert F_N^{SW}(q)\vert^2 + \frac{1}{Z}(1-\vert F_N^{SW}(q)\vert^2)\vert F_p(q)\vert^2
\label{eq:but}
\end{equation}
Equation \ref{eq:but} is the same as Eq. 4 of (\cite{ref:Butkevich})
neglecting the atomic form factor which is only relevant at very small
momentum transfer.  For the proton form factor $F_p$ we have used an
exponential charge density (dipole) with
$\rho(r)=R_p^{-1}e^{-\frac{r}{R_p}}$ and $R_p =0.71$~fm.

\begin{figure}[!t]
\centering
\includegraphics[width=9cm]{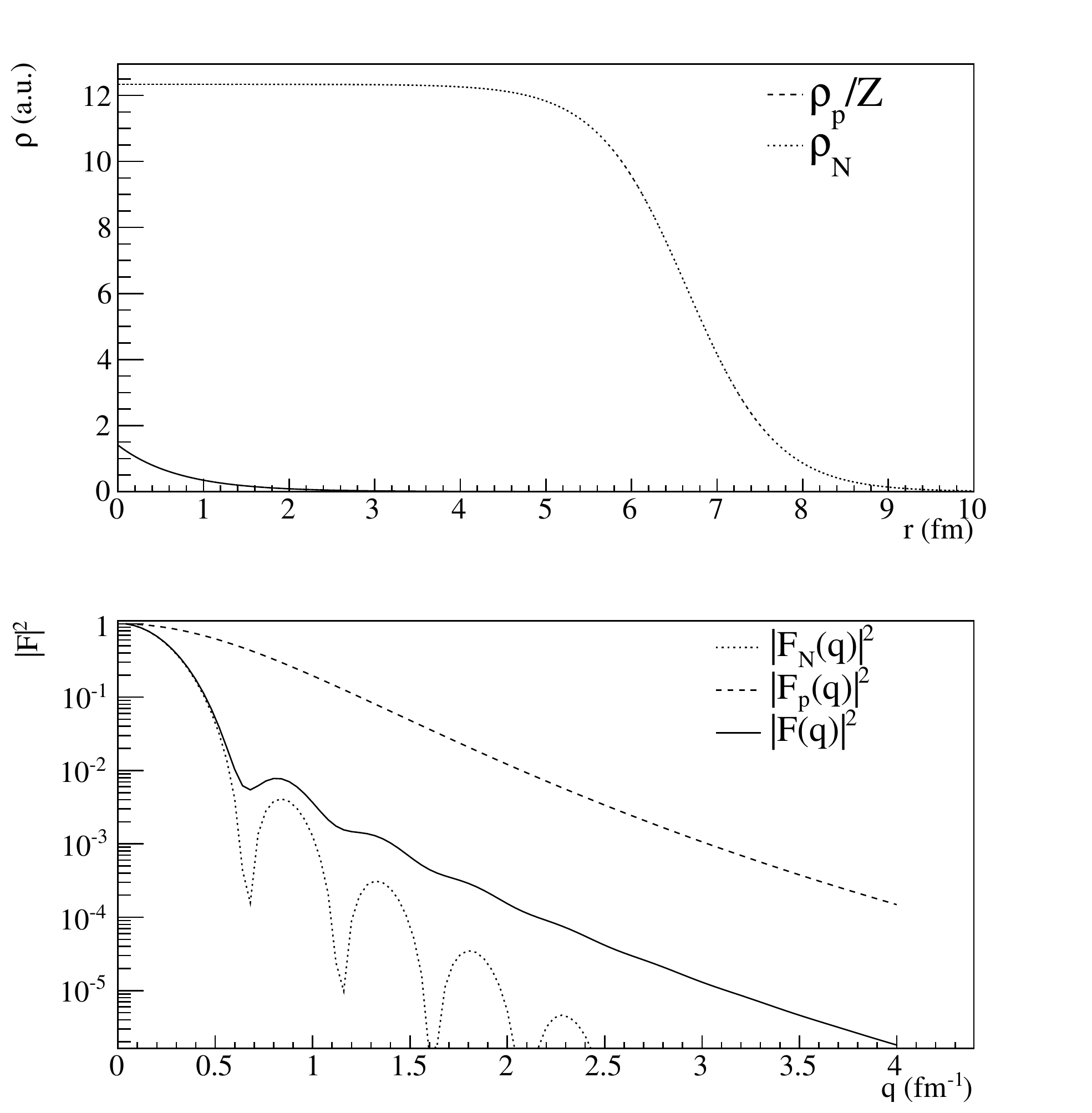}
\caption{Top: charge density for the Lead nucleus according to the
  Saxon-Woods parametrization of Eq.~\ref{eq:SW} with the parameters
  described in the text (dotted line) and for the proton (continuous
  line). The relative size is given by $Z=82$.  Bottom: the nuclear
  form factor obtained by the squared Fourier transform of
  $\rho_{SW}(r)$ ($|F_N(q)|^2$, dotted), the proton form factor
  obtained by the squared Fourier transform of $\rho_p(r)$
  ($|F_p(q)|^2$, dashed). The combined form factor ($|F(q)|^2$,
  continuous) as in Eq.~\ref{eq:but}.}
\label{fig:usedFF}
\end{figure}

The Lead form factor $|F_N^{SW}(q)|^2$ used in the simulation is shown in
Fig.~\ref{fig:usedFF} (solid curve). The inclusion of scattering off
protons (dotted curve) increases significanlty the probability of
large-angle scattering as shown in Fig.~\ref{fig:usedFF}, bottom, as
it can be seen by comparing the solid ($\vert F(q)\vert^2$) and dotted
($\vert F_N(q)\vert^2$) curved. In summary the sampling is applied
using a $g_{rej}^\prime(\theta) = g_{rec}(\theta)\times\vert
F(q)\vert^2\times g_{a}(\theta)$.  At the interesting energies the
effect of atomic corrections is essentially absent ($g_a>1$ for any angle)
and recoil introduces a modest correction due to $p/M$ being $\simeq 0.5-7\%$.

\section{Benchmarking the Monte Carlo with theoretical predictions}
\label{sec:G4theo}
\begin{figure}
\centering
\includegraphics[width=9cm]{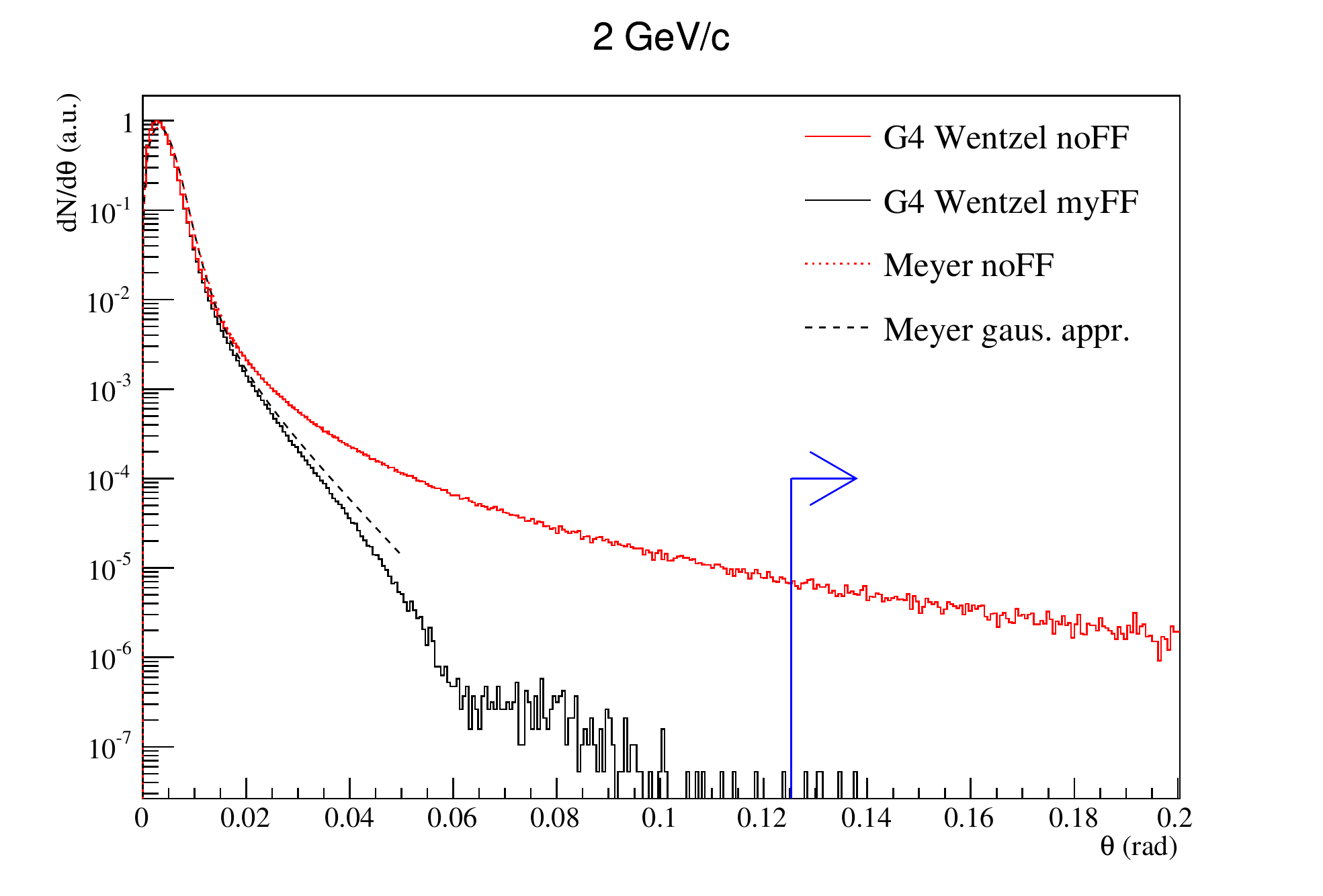}
\caption{Distribution of the scattering angle for 2~GeV/c muons
  impinging orthogonally on a 1~mm thick lead target according to the
  prediction from the theory of Meyer (lines) and to the GEANT4 based
  simulation described in the text (histograms). Both the cases of a
  point-like or extended nucleus are shown in red and black
  respectively. The arrow marks the angle corresponding to the OPERA
  signal region (Sect.\ref{sec:OPERA}).}
\label{fig:meyer2}
\end{figure}

\begin{figure}
\centering
\includegraphics[width=9cm]{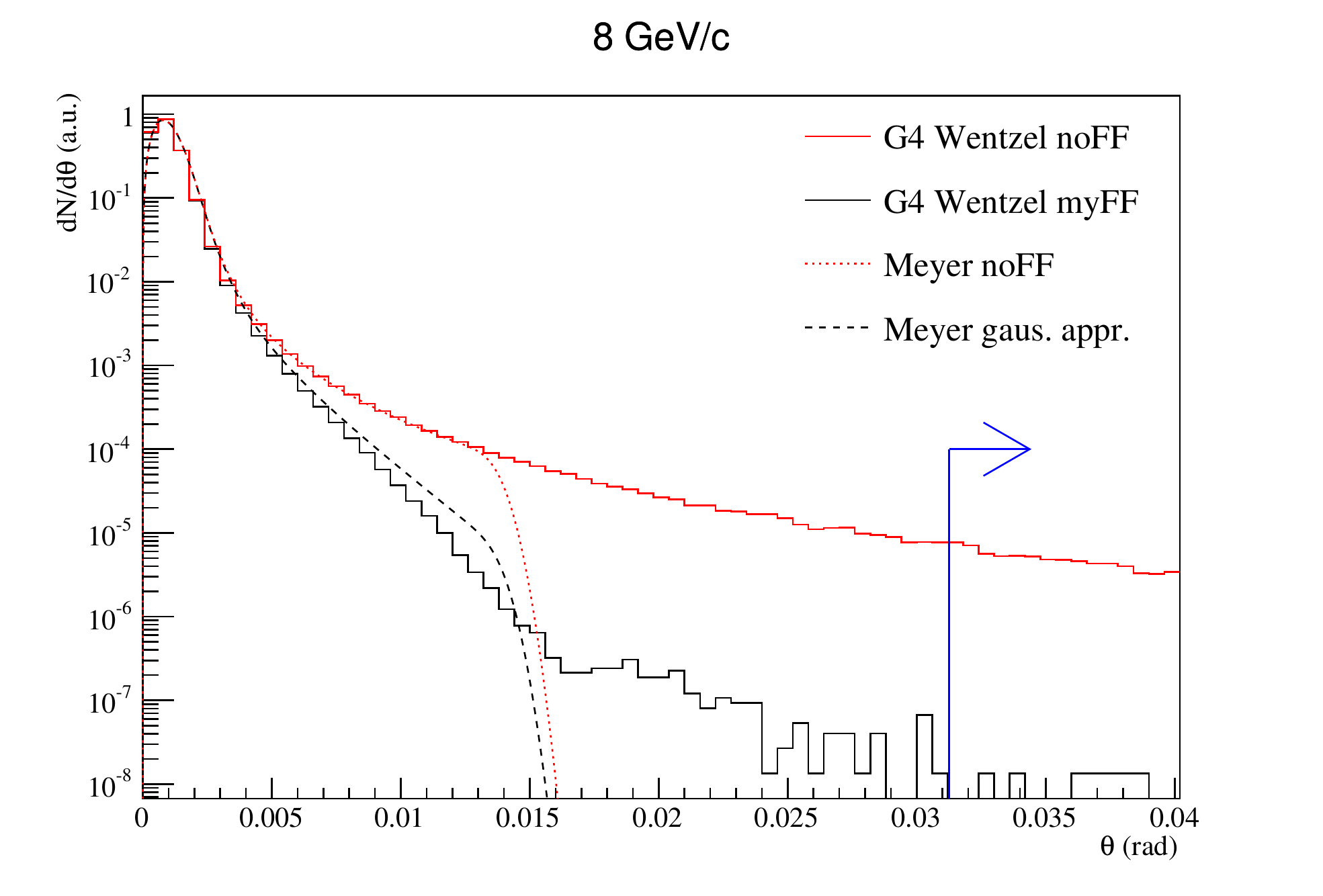}
\caption{As for Fig.~\ref{fig:meyer2} but for 8 GeV/c muons.}
\label{fig:meyer8}
\end{figure}

\begin{figure}
\centering
\includegraphics[width=9cm]{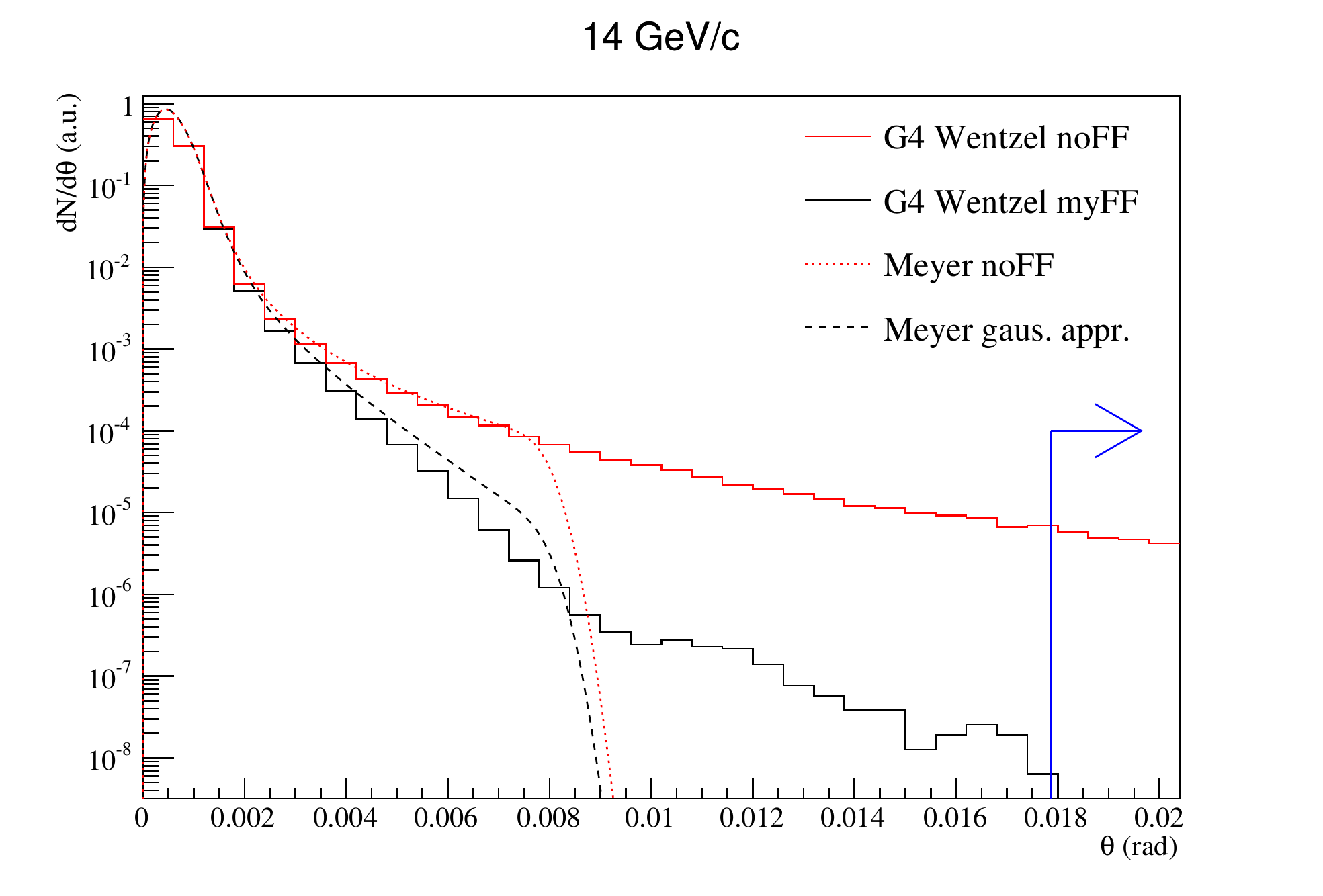}
\caption{As for Fig.~\ref{fig:meyer2} but for 14 GeV/c muons.}
\label{fig:meyer14}
\end{figure}

We have compared the result of the Monte Carlo simulation of
Sect. \ref{sec:implem} for muons of 2, 8 and 14 GeV/c momenta and
orthogonal incidence on a 1 mm thick Lead foil with the predictions of
the theory by Meyer discussed in Sect.~\ref{sec:theory}. The results
are shown in Fig.~\ref{fig:meyer2} (2~GeV/c), Fig.~\ref{fig:meyer8}
(8~GeV/c) and Fig.~\ref{fig:meyer14} (14~GeV/c).  The Monte Carlo
distribution is shown by the histogram while theoretical predictions
are shown by the dashed curves. Both the expectation for a point-like
(red) or diffuse (black) charge distribution are shown. It can be
noticed that the theoretical predictions show a steep drop above a
certain value which depends on the muon momentum. This is understood
in terms of numerical accuracy of the program (the sums in
Eqns.~\ref{eqA} and \ref{eqB} involve terms with large powers of
$\theta$). This behaviour prevents using the Meyer theory for a
quantitative estimate of LAS, as anticipated.  Nevertheless we note
that, despite the approximations used in deriving the theoretical
curves (Sect. \ref{sec:theory}), below the critical maximal angle,
there is a resonable agreement with the Monte Carlo results.


\section{Benchmarking the Monte Carlo with experimental data}
\label{sec:G4data}
We have tested the capability of the modified Monte Carlo
(Sect. \ref{sec:implem}) to reproduce several data-sets available in
the literature. The goal of these experiments was to investigate the
scattering of muons or electrons at high momentum transfer to study
the nuclear charge distribution. They are hence particularly suited
for our purpose despite being rather old. A certain degree of
extrapolation is present with respect to the CNGS case due to the use
of different energy and particles and targets. This is mitigated by
the fact that a significant variation of these parameters is sampled
and that overall the conditions are not too dissimilar to the case of
interest. A summary of the basic parameters of the used
data-set is given in Tab.~\ref{tab:datasets}.
\begin{table}
\centering
\begin{tabular}{|c|c|c|c|c|c|}
\hline
$p$ (GeV/c) & particle &  material & $t$ (mm) & $n_{data}$ & Ref.\\
\hline
7.3  & $\mu$ & Cu & 14.4 & $3.1 \times 10^4$ & \cite{Akimenko}\\
\hline
11.7  
& $\mu$ 
& Cu 
& 14.4 
& $8.7 \times 10^3$ 
& \cite{Akimenko}\\
\hline
2.0 
& $\mu$ 
& Pb 
& 12.6 
&$2.5 \times 10^7$
& \cite{Masek}\\
\hline
0.512 
&$e$
& Pb
& 0.217 
& $\sim (1.25 \times 10^{14})/\rm{s}$ 
& \cite{Frois}\\
\hline
$[$1, 15$]$  
&$\mu$
& Pb 
& 2.0 
& 
${\mathcal{O}}(10^4)$
& \cite{OPERApropA}, \cite{OPERApropB}\\
\hline
\end{tabular}
\caption{Experimental data-sets used for the benchmarking of the simulation. $n_{data}$
indicates the number of measured particles.}
\label{tab:datasets}
\end{table}
The Copper targets are about a factor 3-4 thicker than the OPERA case while the momentum is in the
OPERA window. The Lead data are represented by a target being thinner by about a factor ten (low-momentum
electrons) and one being about a factor five thicker but in the OPERA momentum window. It must
be noted that however the characteristic quantity of the process is the transverse momentum transfer $q$
rather than the incident momentum. The ability of the simulation in describing the data in this space
is a good indicator of its reliability in the OPERA region of interest. 

\subsection{Copper data with muons (Akimenko et al.)}
\label{sec:copper}
The experiment~\cite{Akimenko} was performed in 1986 at the IHEP
accelerator using the HYPERON setup to measure the scattering of muons
of energies of 7.3 and 11.7 GeV/c off a 14.4 mm thick (one $X_0$)
Copper target.  The incoming muon direction was determined with four
proportional chambers allowing a spatial resolution of 0.7 mm and an
angular accuracy of 0.3 mrad.  A magnet was also employed allowing the
incoming momentum to be estimated with a 3\% relative
error. Downstream of the target, the positions and angular errors were
reduced by about a factor 2 w.r.t to the upstream section. Full
efficiency is claimed up to 50 mrad ($q_{max}=1.9$~fm$^{-1}$ for 7.3
GeV/c and 3.0~fm$^{-1}$ for 11.7 GeV/c). A 1.5 m thick iron absorber
was used to stop pions. A selection based on a pair of Cherenkov
counters upstream of the target allowed to achieve a pion
contamination below 2\%. In addition, the requirement of single track
scattering was applied to reject events with nuclear interactions of
the pion contamination.

\begin{figure}
\centering
\includegraphics[width=9.5cm]{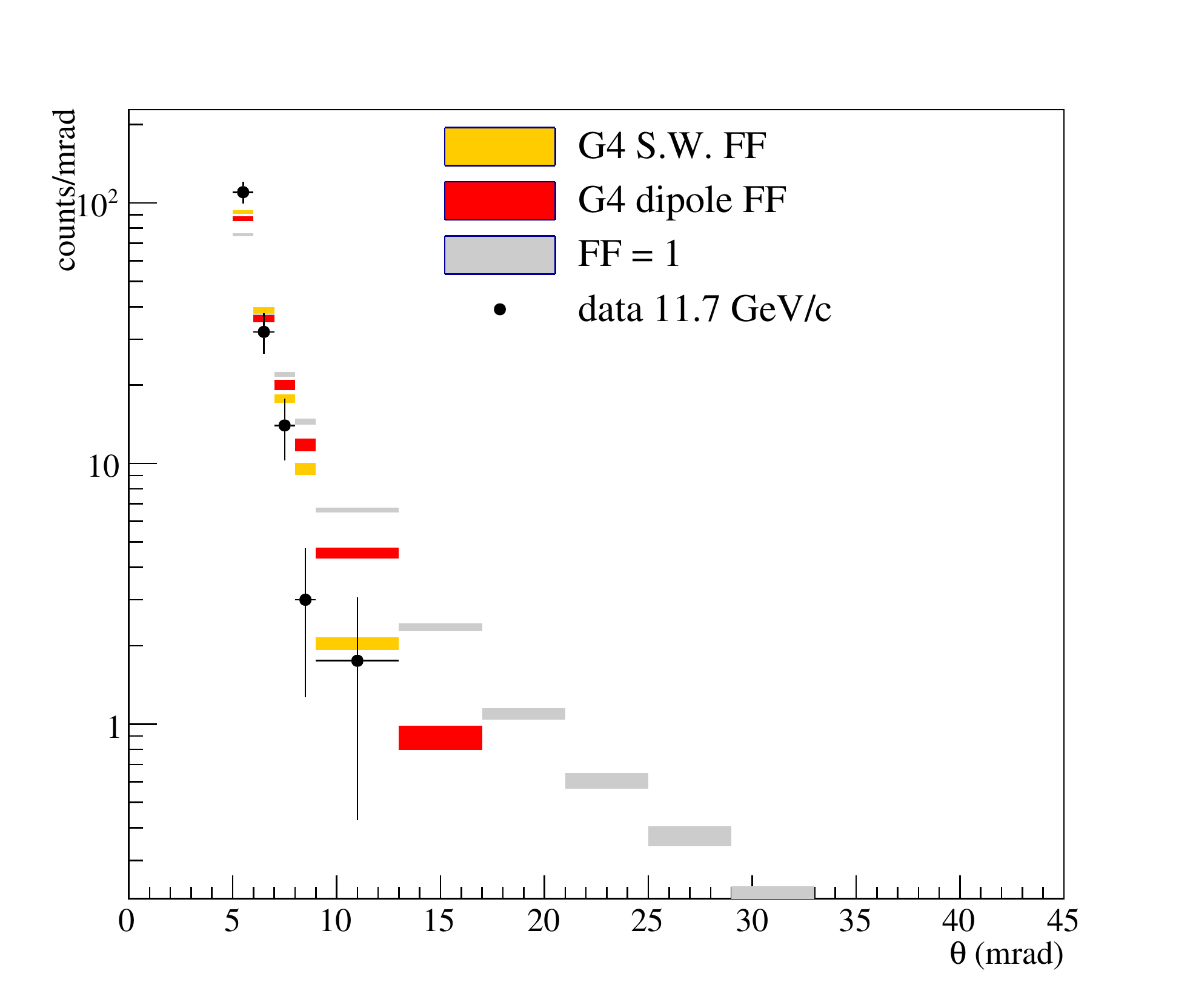}
\caption{Comparison of the Monte Carlo simulation to the 11.7 GeV/c data-set. 
Data have been obtained by digitisation from the original paper\cite{Akimenko}.
\label{fig:Cu1}}
\end{figure}

\begin{figure}
\centering
\includegraphics[width=9.5cm]{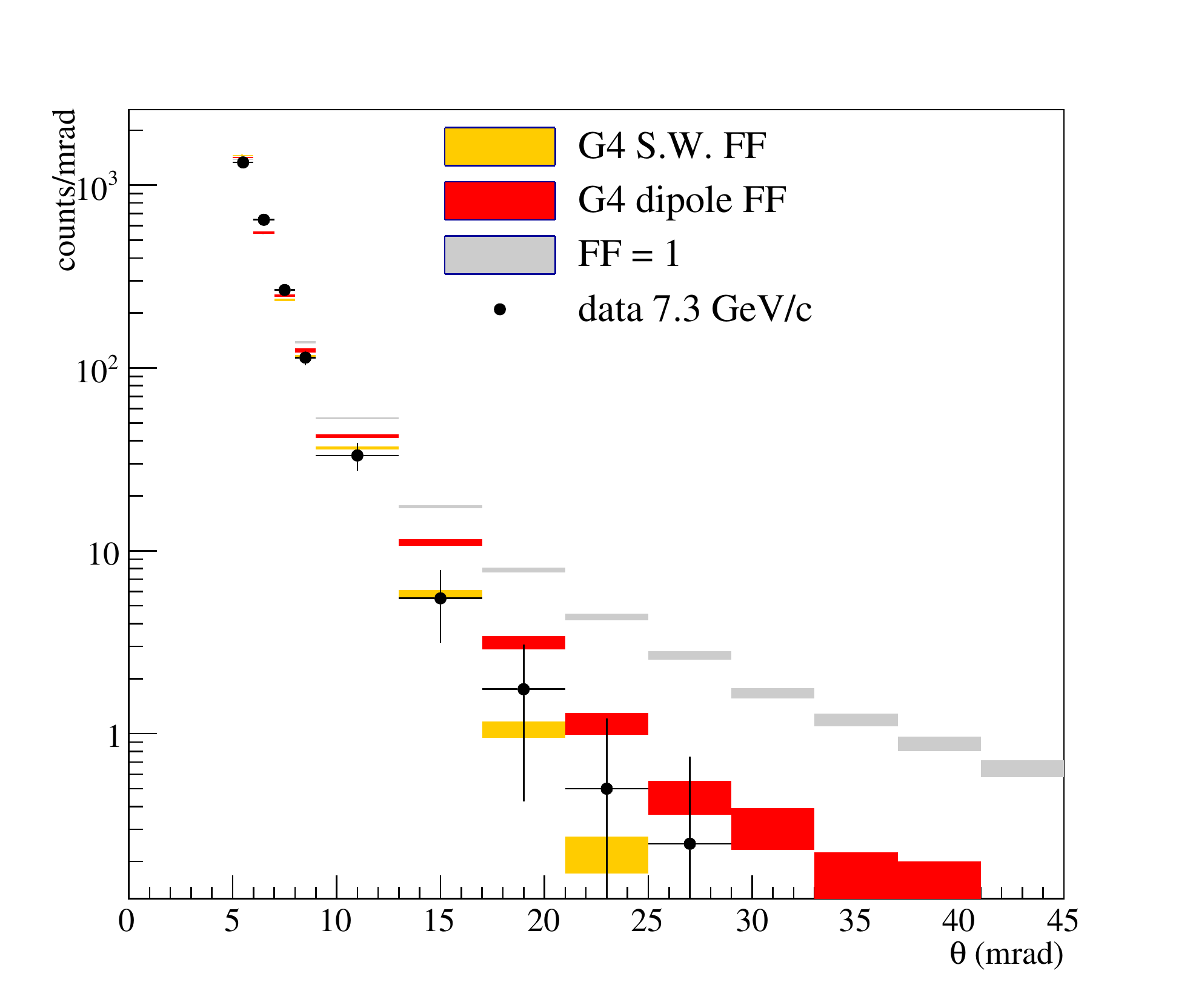}
\caption{Comparison of the Monte Carlo simulation to the 7.3 GeV/c data-set.
Data have been obtained by digitisation from the original paper\cite{Akimenko}.
\label{fig:Cu2}}
\end{figure}

The scattering in the counters was deconvoluted by using data-sets
obtained after the target had been removed.  The comparison with the
simulation is performed for $\theta>5$~mrad since in this region the
background contribution can be neglected.  Real data and Monte Carlo
events are normalised to each other above this threshold. The results
are shown in Fig.~\ref{fig:Cu1} and Fig.~\ref{fig:Cu2} for 11.7 GeV/c
and 7.3 GeV/c muons respectively.  In this case the same procedure
used for Lead was employed except for the fact that the Saxon-Wood
parametrisation of the Copper nucleus charge density was
used. Experimental data are shown by the bullets with error bands. The
Monte Carlo prediction is shown with three options: 1) a point-like
nucleus prediction ($|F(q)|^2=1$) (gray). 2) the dipole nuclear form
factor (red) 3) the improved form-factor of Eq.~\ref{eq:but}
(yellow). In both cases the mixed algorithm described in
\ref{sec:implem} has been used. The band width in Monte Carlo
histograms represent the statistical uncertainty. This meaning of
symbols is also preserved in the following. The disagreement with the
point-like simulation indicates that the nuclear density is in this
case actually probed.  Furthermore a better description is obtained by
using the improved form-factor with respect to the dipole one.

\subsection{Lead data with muons (Masek et al.)}
\label{sec:Masek}
The experiment\cite{Masek} was performed in 1961 profiting of the Bevatron accelerator at
the Lawrence radiation laboratory employing a muon beam with a median
momentum of ($2.00~\pm~0.03$)~GeV/c and a 3.5\% spread.
The total number of muons incident on the apparatus was $2.5\times
10^7$.  The lead target had a thickness of 14.4~g/cm$^2$ corresponding
to 1.268~cm (one half inch). Scattered particles were observed up to
12$^\circ$ ($p_T~\simeq~400$~MeV/c). The muon beam is obtained from a
$(3.5\pm 0.3)$ GeV/c pion beam with magnetic selection (see Fig.~3 of
the original paper). 2~GeV/c muons are those decaying backward in the
$\pi$ rest frame.  This large difference in momentum allows a clean
$\pi/\mu$ separation.  The pion contamination on the target is
estimated to be of about 3\%.  By using a 107 cm (42'') thick iron
absorber downstream of the target and an upstream Cherenkov counter
the effective pion contamination was reduced to $4.9 \times
10^{-6}$. Emulsion detectors downstream of the target were also
employed to control the pion contamination.

The incoming and outgoing muon directions were determined by using
four identical counter hodoscopes made of scintillator bars at
positions $A_A$, $A_B$, $A_C$ and $A_D$ (Fig. 3 of \cite{Masek}).  The
acceptance of such an arrangement was between 2 and 14$^\circ$. Each
station was composed of 20 vertical scintillators each having
dimensions (0.95, 2.54, 15.24)~cm, the smallest dimension being the
one in the horizontal plane. The downstream stations centers were
shifted away from the beam direction by 26.2~cm. The distance along
the beam between the stations was 2.794 m (AB) and 2.007 m (CD).

The beam spread in the horizontal and vertical directions was
determined triggering with a small scintillator ($S$) located along
the beam axis in between stations $A_C$ and $A_D$, rotating the $A_A$
and $A_B$ stations by 90$^\circ$. The coincidence rate between strips
in scintillators in $A_A$ and $A_B$ is shown in Fig.~6 of the original
paper.  This bidimensional rate map was used to generate the incoming
direction of muons in the simulation accordingly. The vertical spread
was generated assuming a gaussian distribution with a $\pm 0.6^\circ$
r.m.s. This is supported by the paper which states a spread of $\pm
0.8^\circ$ and $\pm 0.6^\circ$ in the horizontal and vertical views
respectively.  A gaussian fit of the $\theta_x$ distribution obtained
from the information contained in the hit strip bidimensional
distribution (Fig.~6 of \cite{Masek}) and the detector geometry gives
a spread of about 0.6$^\circ$ not too far from the 0.8$^\circ$ quoted
in the paper.

We have reproduced with a simple simulation both the angular spread of the
beam and the finite-solid angle effects introduced by the strip width.
This was performed by binning the true positions according to the
detector geometry described above. A simple linear trajectory has
been assumed in between different scintillator stations.

\begin{figure}
\centering
\includegraphics[width=9.5cm]{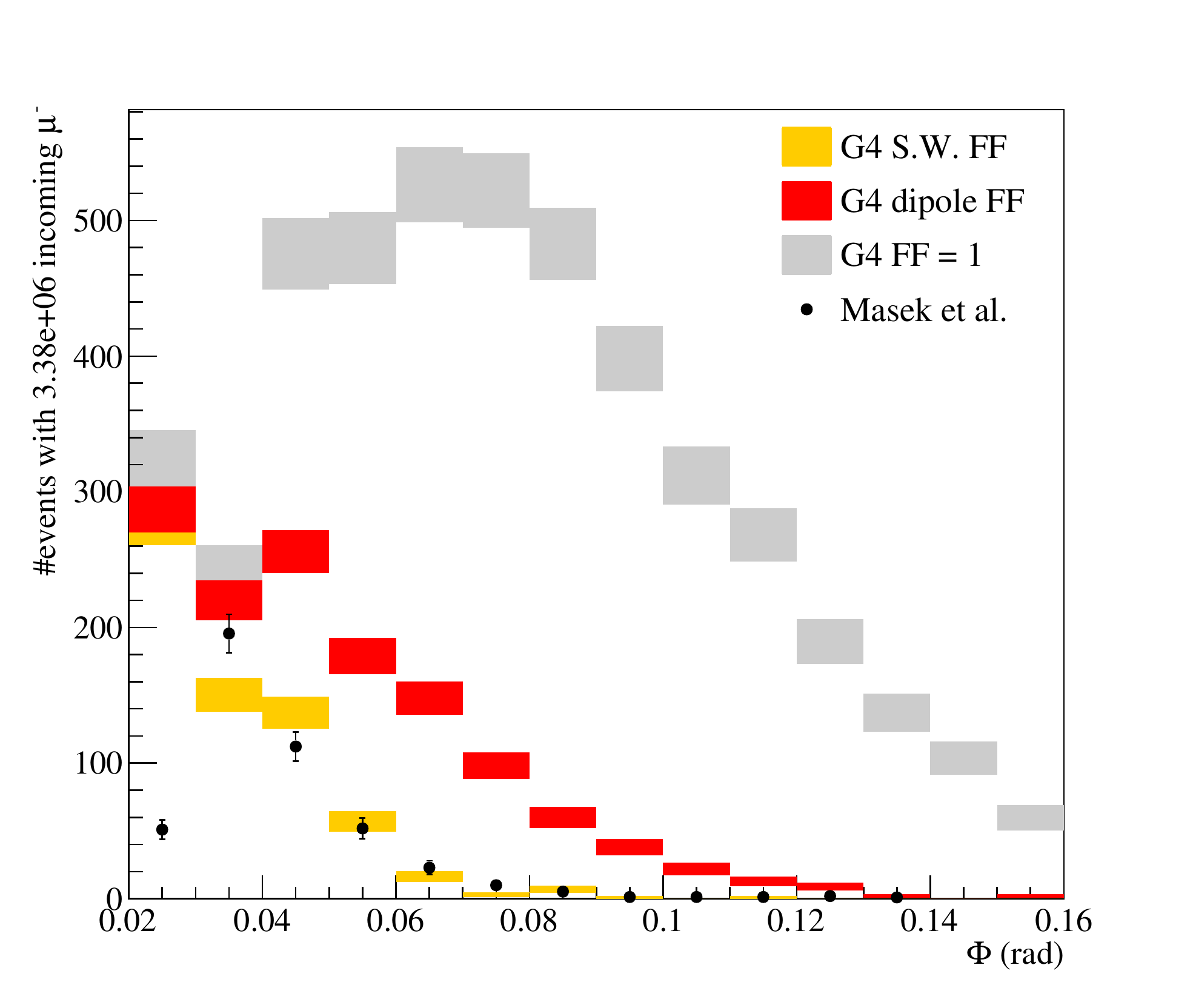}
\caption{Comparison of the Monte Carlo simulation to the Masek et al. data-set.\label{fig:Masekdata}}
\end{figure}

The measured quantity is the angle $\Phi$ which is obtained
considering the layout of hit strips in the scintillator counters
which are only measuring the horizontal projection of the scattering
angle (single-view readout). The distribution of $\Phi$ is shown in
real data (bullets) and Monte Carlo (histograms) in
Fig.~\ref{fig:Masekdata}.  The meaning of the symbols is the same as
in Figs.~\ref{fig:Cu1} and \ref{fig:Cu2}. The Monte Carlo
normalisation corresponds to the same number of incoming muons as in
the data (Tab.~\ref{tab:datasets}). In this case both the point-like
and the dipole form-factor simulations fail to describe the
measurement while the modified simulation is capable of reproducing
the experimental points at large scattering angles. The marked
disagreement in the first bin is supposedly caused by the simplified
simulation which might lack some of the geometrical details of the
real apparatus.

\subsection{Lead data with electrons (Frois et al.)}
\begin{figure}
\centering
\includegraphics[width=9.5cm]{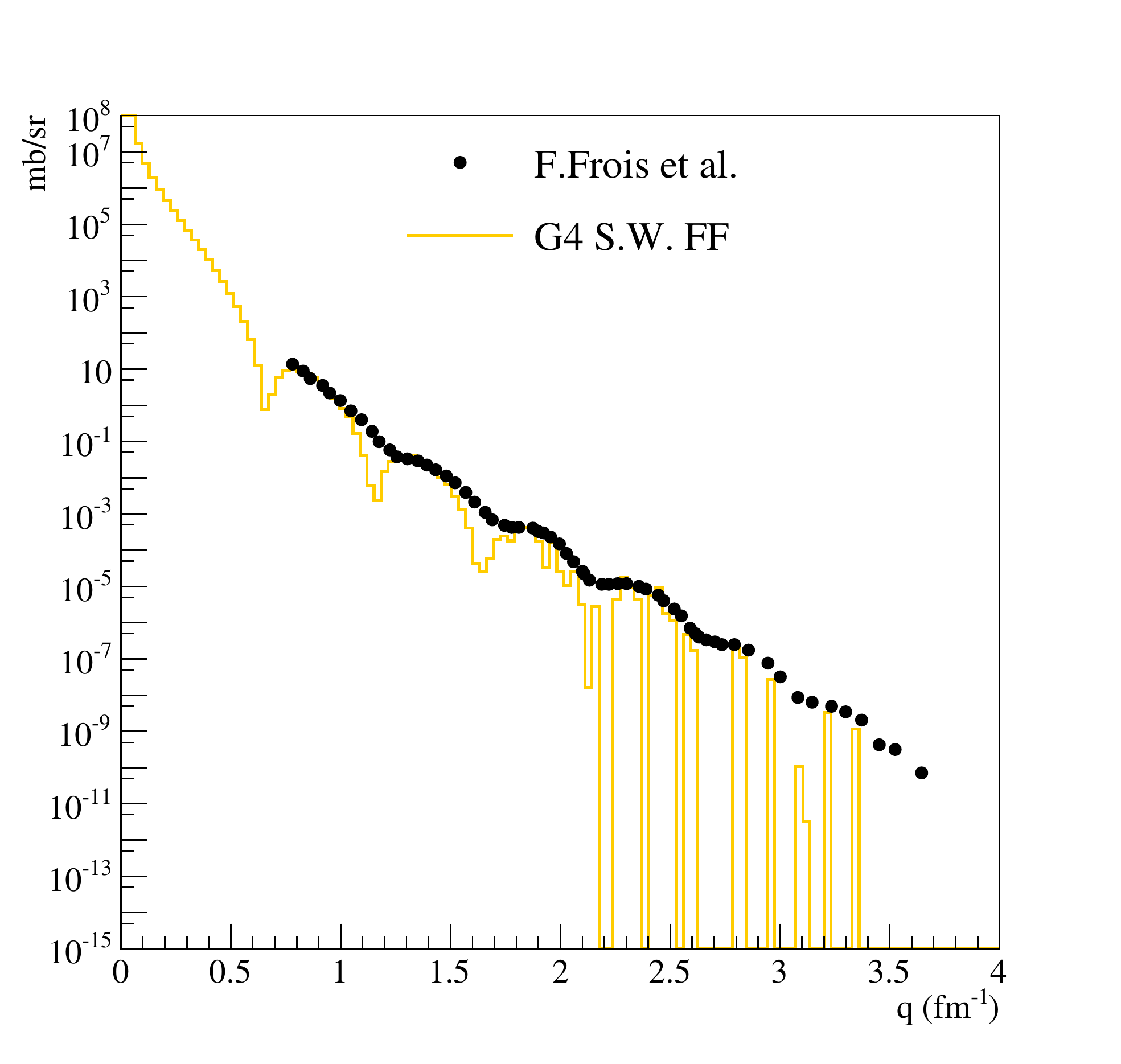}
\caption{Comparison of the Monte Carlo simulation to Frois et al. data-set.}
\label{fig:Saclay}
\end{figure}

The experiment \cite{Frois} was performed in 1977 at Saclay directing
an electron beam with an energy of 502 MeV and a 20 $\mu$A current
($\sim 1.25 \times 10^{14}~e$/s) on a water cooled $217 \pm 2$
mg/cm$^2$ thick lead target.  Thanks to the large event sample the
data allowed a precision on the charge density at the lowest values of
$r$ ever reached at those times or equivalently they allowed to
determine the nuclear form factor at momentum transfers $q$ up to 3.5
fm$^{-1}$ (cross sections down to $10^{-10}$~mb/sr). The measured
distribution of the momentum transfer $q$ spans over 12 decades
(Fig.~\ref{fig:Saclay}, bullets). The scattering angles were
determined with an accuracy of 0.05$^\circ$. The experimental points
are compared in Fig.~\ref{fig:Saclay} with our simulation.  In this
case an event-by-event weighting was applied to the point-like
distribution following the form-factor of Eq.~\ref{eq:but} using the
integral momentum transverse in the target as reweighing
variable. This was dictated by the difficulty in adopting the
single-scattering sampling approach within a reasonable amount of CPU
time. It was verified anyway that the sampling approach correctly
describes the experimental distribution up to $q
\simeq$~1~fm$^{-1}$. In the high-$q$ region 
the two approaches are expected to be
very similar due to the prominence of single scatterings. The
sharpness of the dips in the simulation is an artifact of the
event-by-event instead of scattering-by-scattering form-factor
reweighting.

\section{Large-angle muon scattering in the OPERA experiment}
\label{sec:OPERA}
\begin{figure}
\centering \includegraphics[width=9cm]{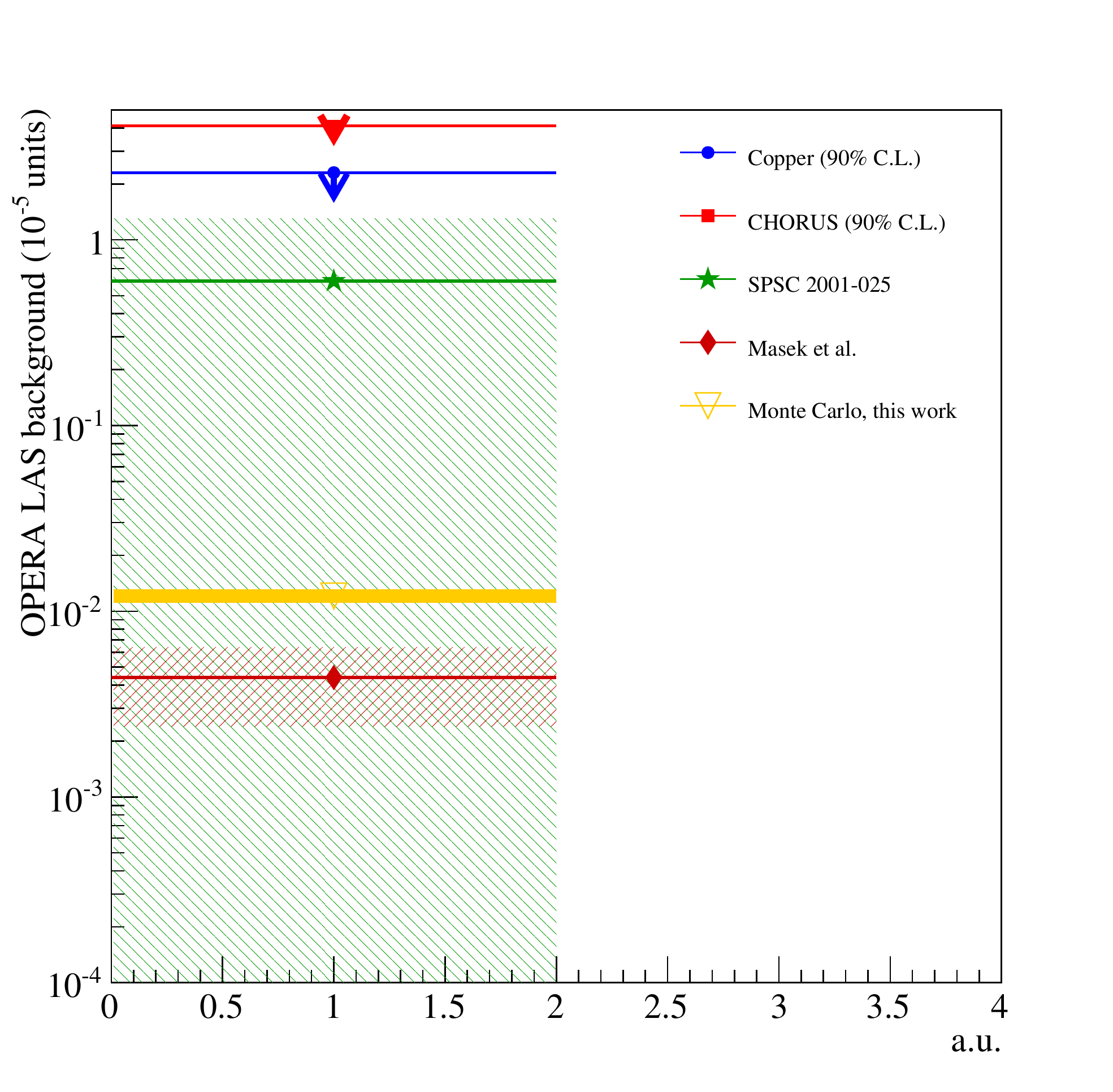}
\caption{A compilation of experimental results on Large
  Angle muon Scattering. Bands represent $1\sigma$ confidence
  limits. The plot is discussed in the text.}
\label{fig:las}
\end{figure}

The large-angle scattering of muons from $\nu_\mu$ CC interactions
represents a background for the $\tau \to \mu$ channel in the OPERA
experiment at the CNGS neutrino beam. The kinematic selection
described in the experiment proposal~\cite{OPERApropA},
\cite{OPERApropB} defines the signal region for the $\tau \to \mu$ 
search by requiring:
\begin{itemize}
\item $p_T^\mu>0.25$~GeV/c
\item $\theta^\mu>20$~mrad
\item $p_\mu>1$ and $p_\mu<15$~GeV/c
\end{itemize}
Furthermore the kink is asked to lie within the first two
lead-emulsion layers downstream of the primary interaction vertex.
One $\nu_\tau$ candidate in this decay channel was
observed~\cite{ref:tau3} with $p_T^\mu=(0.69 \pm 0.05)$~GeV/c,
$\theta^\mu = (245 \pm 5)$~mrad and $p_\mu= (2.8 \pm 0.2)$~GeV/c.  
In the experiment proposal~\cite{OPERApropA,OPERApropB} some data-driven estimates of
this process are reported. They are summarised in Fig.~\ref{fig:las}.
The red arrow represents an upper limit from the non-observation of
high-angle scattering in $1.5\times10^6$~mm of muon tracks in the
nuclear emulsions of the CHORUS experiment. This result was translated
to 2~mm of Lead with the CNGS spectrum yielding~\cite{OPERApropA} a
90\%~CL limit at $4.1\times 10^{-5}$.  The blue line corresponds to
the already described measurement on a Copper target
(Sect.~\ref{sec:copper}). A limit of $2.3 \times 10^{-5}$ at 90\% CL
was obtained by extrapolating the observation to the case of
scattering on Lead assuming a scaling going as $Z^2 \rho / Ap^2$.  The
green hatched band was obtained from the observation of a single
high-scattering event in the preliminary phase of a dedicated
test-beam experiment~\cite{OPERApropB} giving
$0.6^{+0.7}_{-0.6}\times10^{-5}$.  In the proposal and later OPERA
analyses the background has been finally assumed to be
$1.0\times10^{-5}/\nu_\mu^{CC}$ with a 50\% relative error.

\begin{figure}
\centering
\includegraphics[width=9cm]{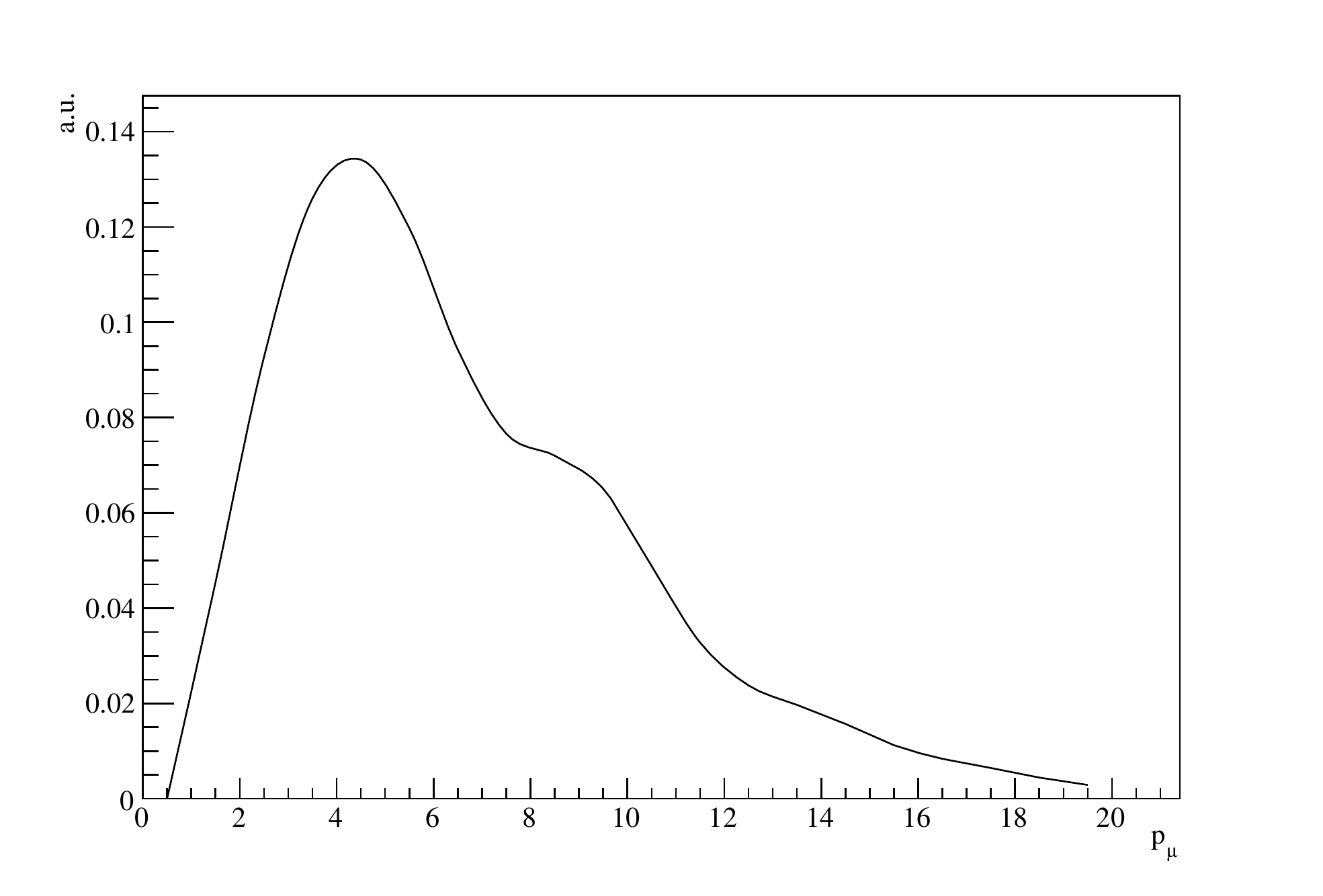}
\caption{Spectrum of $\nu_\mu^{CC}$ CNGS events after selection used for reweighting the
uniformly distributed momentum distribution used as an input to the simulation.
\label{fig:spectrum}}
\end{figure}

The contribution of LAS in this kinematic region was estimated using
the simulation described in~\ref{sec:implem}.  A total number of about
1.1 billion incident $\mu^-$ have been generated with a flat momentum
distribution between 1 and 15 GeV/c with orthogonal
incidence\footnote{The angular distribution of $\nu_\mu^{CC}$ muons is
  such that the average path in lead in reality is larger than what
  assumed here by a few \%.}  on the lead-film double cell. The
obtained distributions have then been reweighted in the incident muon
momentum in order to reproduce the real spectrum of CNGS
$\nu_\mu^{CC}$ events.  The used reweighting
function~\cite{ref:spectrum} (Fig.~\ref{fig:spectrum}) reflects the
expectation after applying the full selection chain used in the
analysis~\cite{ref:2taupaper}.

It must be noted that in the real case of the OPERA experiment the
momentum is determined with a certain resolution and that the
selection cut $p_T>250$~MeV/c has to be taken on the reconstructed
variable. The momentum resolution of muon tracks depends on whether
the muon stops in the target or it exits from it. Furthermore if the
track does not cross the spectrometer the only available determination
comes from the analysis of multiple Coulomb scattering in the emulsion
detectors.  A very conservative estimate of the achievable resolution
has been taken by considering a 15\% Gaussian smearing in the $p^{-1}$
variable (See f.e. Fig.~91 of~\cite{OPERApropA}).

\begin{figure}
\centering
\includegraphics[width=8cm]{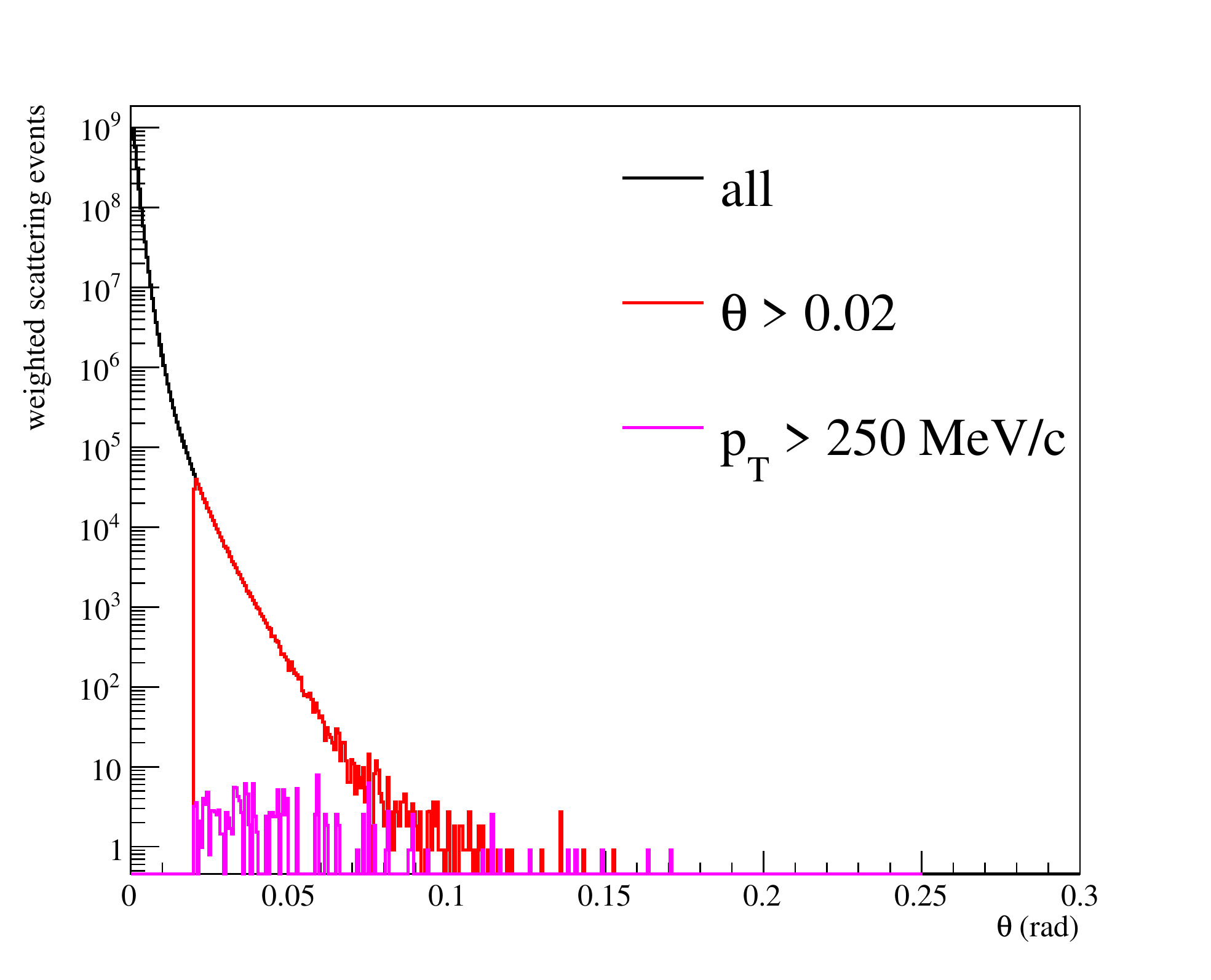}\\
\includegraphics[width=8cm]{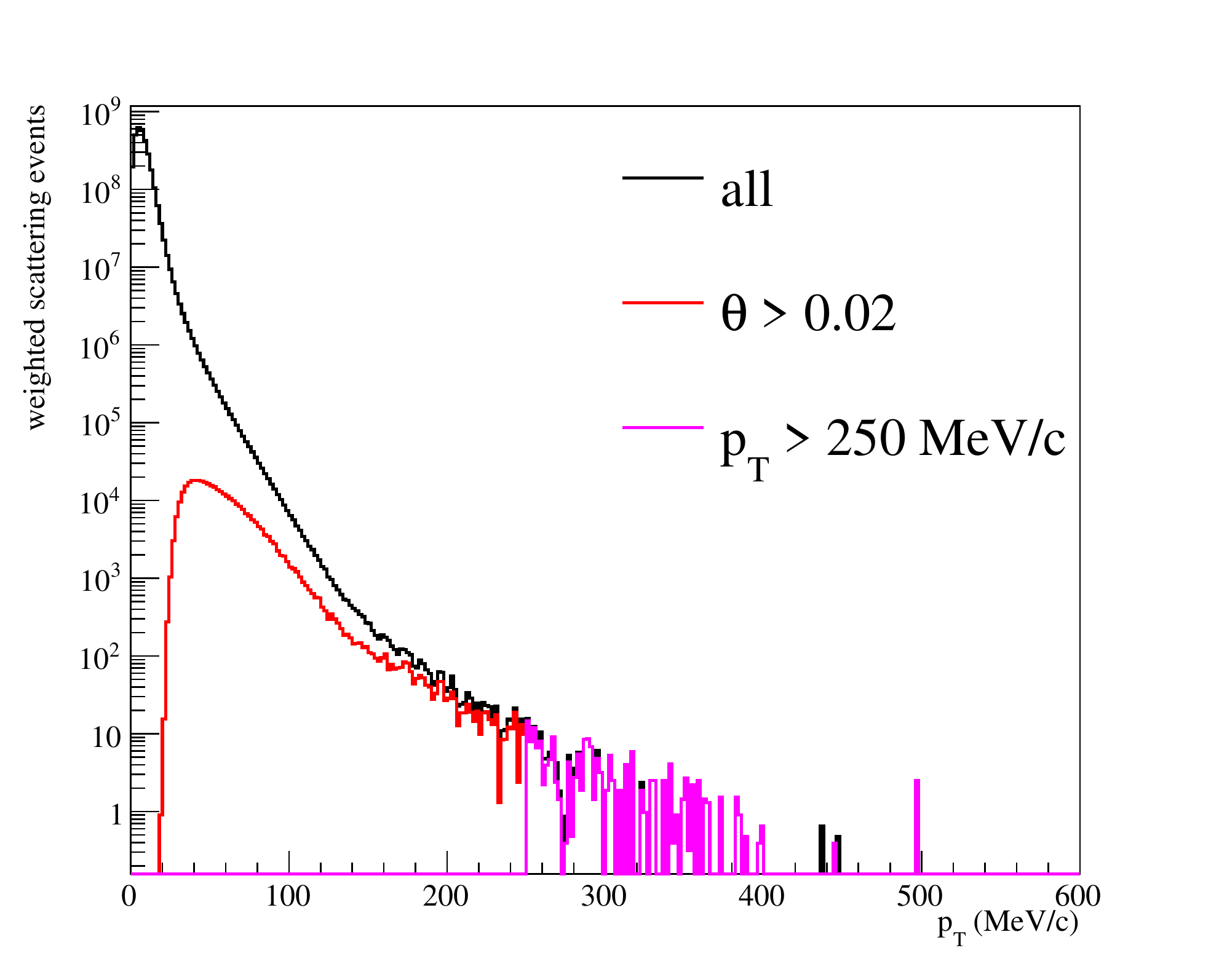}
\caption{Distributions of $p_T^\mu$ (top) and $\theta_\mu$ (bottom) before cuts (black), after the angular cut (red)
and after the angular and $p_T$ cuts (magenta).}
\label{fig:distrib1}
\end{figure}

The scattering angle $\theta_\mu$ is defined as the difference in
slope at the entrance and exit of the lead foil.  We show in
Fig.~\ref{fig:distrib1} the distributions of $\theta_\mu$ (top) and
$p_T^\mu$ (bottom) before cuts (black), after the angular cut (red)
and after the angular and $p_T$ cut (magenta).  The numbers are break
down as the cuts are applied in Tab.~\ref{tab:resG4}.
\begin{table}
\centering
\begin{tabular}{|c|c|c|c|}
\hline
& crossed cells& with $\theta>20$~mrad& and $p_T>0.25$ GeV/c\\
\hline
\hline
n. &2.22952 $\times 10^9$ & 1082520 & 115\\
\hline
weighted &3.05367 $\times 10^9$ & 365195 & 179.155 \\
\hline
fraction & 1 & $2.4 \times 10^{-4}/\nu_\mu^{CC}$& $(1.2 \pm 0.1) \times 10^{-7}/\nu_\mu^{CC}$ \\
\hline
\end{tabular}
\caption{Simulation results for the OPERA LAS background estimate including momentum 
reconstruction uncertainties. \label{tab:resG4}}
\end{table}
The probability of having a LAS event in the signal region over two mm of Lead is then found to be:
$f_{LAS}^{Pb}= (1.2 \pm 0.1\rm{(stat.)})\times 10^{-7}/\nu_\mu^{CC}$ (yellow band in Fig.~\ref{fig:las}).
The effect of momentum smearing convoluted with the observed
distributions has a sensible effect since it increases the background
due to migration of events in the high-$p_T$ region by bringing it
from to $(6.8 \pm 0.8) \times 10^{-8}/ \nu_\mu^{CC}$ to $(1.2 \pm 0.1)
\times 10^{-7}/ \nu_\mu^{CC}$.

A rough estimate of the OPERA LAS background can be also obtained by
considering the data from Masek et al.\cite{Masek} discussed in
Sect.~\ref{sec:Masek}. In that case the muon momentum is 2~GeV/c which
is included in the momentum range considered for the OPERA signal
region ([1, 15]~GeV/c). The thickness is 12.6~mm to be compared with
the 2~mm of OPERA. The projected angle $\Phi$ is considered in that
case. Muons with $p_T > 250$~MeV/c will have approximately $\Phi >
0.088$.  On a sample of $2.5 \times 10^7$ muons from the published
binned distribution we can estimate the observation of (7~$\pm$~3) events.
Scaling linearly for the different thickness one can then estimate:
$f_{LAS}^{Pb,Masek}\rm{(2~GeV/c)} = (0.44 \pm 0.20)\times
10^{-7}/\nu_\mu^{CC}$ which lies even below the Monte Carlo result
averaged in the [1, 15]~GeV/c window (dark red band in Fig.~\ref{fig:las}).

\subsection{Contribution of the scattering in other materials}
The decay of the $\tau$ lepton is also considered when occurring in
the emulsion films. For this reason we have also evaluated the LAS
probability in these materials.  Due to the high number of atomic
components the implementation of Saxon-Woods nuclear form factors was
not attempted in this case. For this reason we can take the results as
an upper limit of the background.  It must be also noted that for
lighter nuclei the use of dipole form-factors, while remaining
necessary, is less compelling than for the case of Lead.  Each film is
composed by a TAC (TriAcetylCellulose, C$_{28}$H$_{38}$O$_{19}$) 200
$\mu$m thick plastic base and by a pair of 50 $\mu$m thick emulsion
layers on both sides of the base.  The TAC has been modeled as a
mixture of Carbon, Hydrogen and Oxygen in the proportion of 28:38:19.
The detailed element budget for the emulsion layers \cite{ref:emucomp}
in terms of mass fraction is Ag 38.34\%, Br 27.86\%, C 13.00\%, O
12.43\%, N 4.81\%, H 2.40\%, I 0.81\%, S 0.09\%, Si 0.08\%, Na 0.08\%,
K 0.05\%, Sr 0.02\%, Ba 0.01\%. The results are: $f_{LAS}^{emul} < 1.2
\times 10^{-7}/\nu_\mu^{CC}$ and $f_{LAS}^{TAC} < 0.7 \times
10^{-7}/\nu_\mu^{CC}$.

\section{Muon photo-nuclear interactions in the OPERA experiment}
\label{sec:photonuc}
For completeness we also try to estimate a possible contribution to
the $\tau \to \mu$ background allowing for the possibility that it
might arise from muon photo-nuclear interaction with an undetected
hadronic remnant\footnote{We have not considered in detail the possible
  contribuition from muon hard bremsstrahlung since the high energy
  gamma associated to the large scattering of the muon can be vetoed
  very efficiently in the emulsion detectors~\cite{emudetEM}.}.
At high enough muon energies ($E>10$~GeV), and at
relatively high energy transfers e.g. when the energy lost by the muon,
$\nu=E^\prime-E$, is more than a few \% of its initial energy, the contribution
from the inelastic interaction of muons with
nuclei~\cite{ref:munuclear} starts to play a non negligible role. 

The differential cross section $\sigma(E,\nu)$ (cm$^{2}$~g$^{-1}$~GeV$^{-1}$)
can be factorised as $\sigma(E,\nu) = \Psi(\nu)\Phi(E,\nu)$~\cite{ref:munuclear1}.
The first function, $\Psi$, reads:
\begin{equation}
\Psi(\nu)=\frac{\alpha}{\pi\nu}N_A\frac{A_{eff}}{A}\sigma_{\gamma N}
\end{equation}
with $\sigma_{\gamma N}=(49.2+11.1\ln
\nu+\frac{151.8}{\sqrt{\nu}})\times 10^{-30}$~cm$^2$ with $\nu$ in GeV~\cite{ref:munuclear5}. The nuclear shadowing
effect~\cite{ref:munuclear4} is parametrised as
$A_{eff}=0.22A+0.78A^{0.89}$ following~\cite{ref:munuclear4} (0.65 for
Lead).
The adimensional function $\Phi$ depends on the fraction $v=E/\nu$ and $E$ as follows:
\begin{equation}
\begin{split}
\Phi(E,\nu)=v-1+\left[1-v+\frac{v^2}{2}
  \left(1+\frac{2\mu^2}{\Lambda^2}\right)\right]\times\\ \ln\frac{\frac{E^2\left(1-v\right)}{\mu^2}\left(1+\frac{\mu^2
    v^2}{\Lambda^2\left(1-v\right)}\right)}{1+\frac{Ev}{\Lambda}
  \left(1+\frac{\Lambda}{2M}+\frac{Ev}{\Lambda}\right)}
\end{split}
\end{equation}
$M$ being the nucleon mass, $\mu$ the muon mass and $\Lambda^2=0.4$ GeV$^2$. 

\begin{figure}
\centering
\includegraphics[width=9cm]{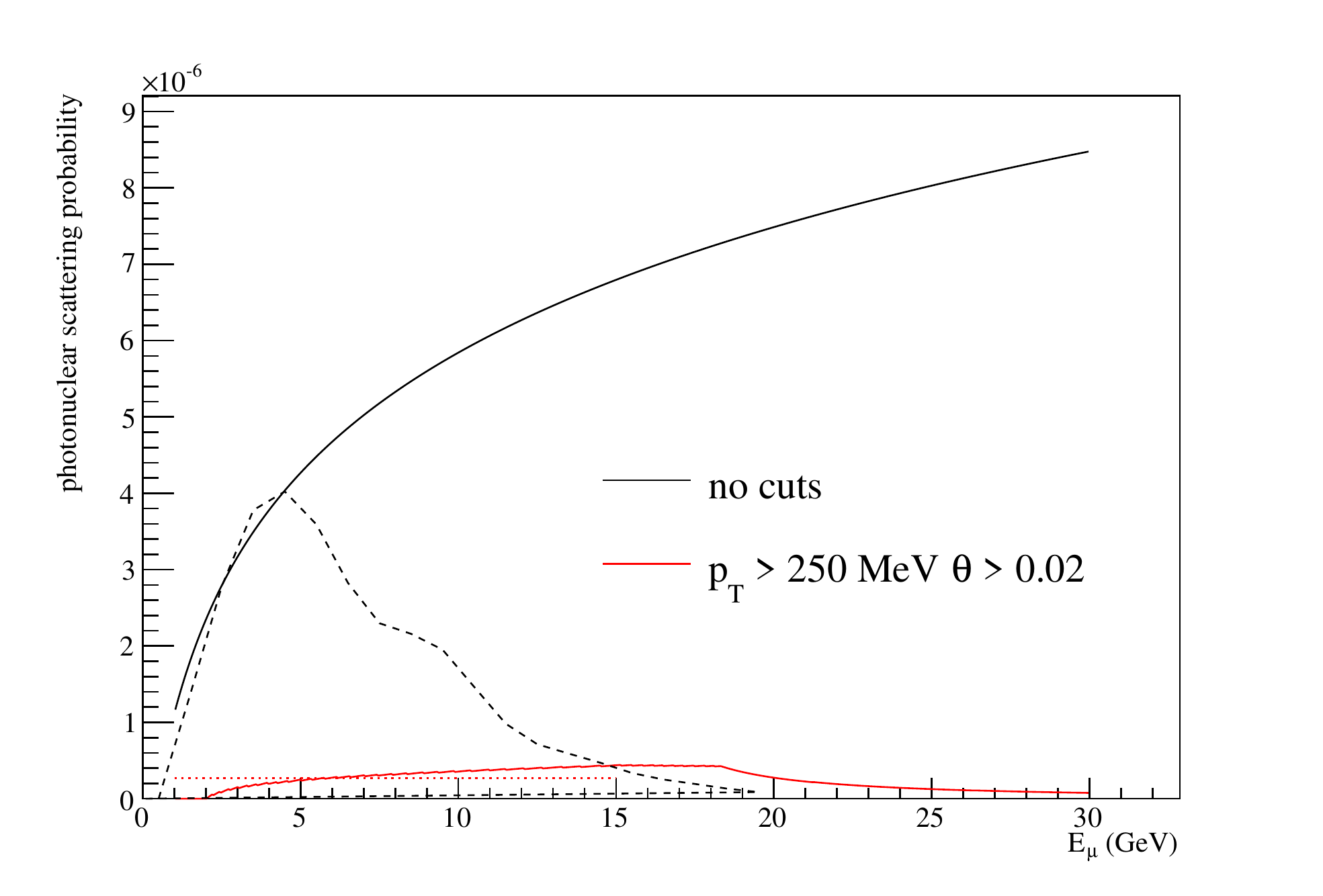}
\caption{Photonuclear probability vs $E_\mu$.}
\label{fig:photonuc1}
\end{figure}

An energy-dependent scattering probability per incoming muon, $P_{\mu
  N}(E)$, is obtained after integration on $\nu$ and multiplication by
$\rho t = 1.135$~g cm$^{-2}$: $P_{\mu N}(E)= \rho
t\int_{\nu_{min}}^{\nu_{max}}\Phi(E,\nu)d\nu$ with $\nu_{min}=0.2$ GeV
and $\nu_{max}=E-M/2$~\cite{ref:GEANT4usermanual}. The result is shown
by the black line curve in Fig.~\ref{fig:photonuc1}. In the same
figure we show with the red line the cross section when one restricts
to the phase space which is relevant for the OPERA background: the
energy\footnote{We approximate momentum to energy here.} of the
outgoing muon ($E^\prime = E-\nu$) should be $1<E^\prime<15$ GeV, the
scattering angle ($\sin \theta = \frac{(E-E^\prime)^2}{4(E
  E^\prime-\mu^2)}$) should be $\theta>0.02$ rad and $E^\prime
\sin\theta >0.25$ GeV.  The result is then weighted with the
probability density function of muons from $\nu_\mu^{CC}$ of CNGS muon
neutrinos (Fig.~\ref{fig:spectrum} and dashed curve on
Fig.~\ref{fig:photonuc1}) giving $f_{\mu N} = 2.7
\times10^{-7}~{\rm{events}}/\nu_\mu^{CC}$ (also shown as a red dotted
horizontal line in Fig.~\ref{fig:photonuc1}).
It must be noted that this has to be considered as an upper limit to
the real background level since a complete inefficiency in the
detection of the hadronic system is assumed here for simplicity.  Here
we refrain from trying an estimate of the minimal hadronic activity
which could be detected in the OPERA emulsion detectors since the
contribution is already small enough for practical purposes.

\section{Conclusions}
The probability for the occurrence of large-angle scattering for
multi-GeV muons in 1~mm thick lead targets has been addressed using a
mixed-approach simulation program based on the GEANT4 libraries.  A
realistic parametrisation of the Lead nuclear density has been
implemented (Saxon-Woods) and scattering off single protons has been
considered.  The developed algorithm has been validated by means of
theoretical considerations and supported by experimental data from the
literature.  The simulation setup predicts a background for the OPERA
$\nu_\mu\to\nu_\tau$ analysis in the $\tau \to \mu$ signal region of
$f_{LAS}^{Pb}= (1.2 \pm 0.1)\times 10^{-7}/\nu_\mu^{CC}$ which is well
below the values which have been considered so far.  Finally we also
calculated, in the CNGS context, an upper limit on background
contributions from the muon photo-nuclear process which might in
principle also produce a large-angle muon scattering signature in the
detector: $f_{\mu N}<2.7 \times10^{-7}~{\rm{events}}/\nu_\mu^{CC}$.

\section*{Acknowledgments}
The authors would like to thank the colleagues from the OPERA
collaboration for encouraging this study and for their fruitful
suggestions.  A. Rohatgi is acknowledged as author of the
WebPlotDigitizer application\footnote{http://arohatgi.info/WebPlotDigitizer} which was
used to acquire experimental data-points from old papers.

\appendices
\ifCLASSOPTIONcaptionsoff
  \newpage
\fi

\section{Moli\`ere theory with nuclear effects (Meyer)}
Given a target with thickness $t$, density $\rho$, atomic weight $A$,
charge $Ze$ and an incoming particle of momentum $p$ with charge $ze$
we can define a ``characteristic angle'', $\chi_c$, as
\begin{equation}
\chi_c = \frac{zZe^2}{p\beta c}\sqrt{\frac{4\pi N_A t \rho}{A}},
\end{equation}
$N_A$ being Avogadro's number.
The ``screening angle'' ($\chi_a$) is defined from the De Broglie wave length of
the incoming particle ($\lambda$) and the Thomas-Fermi radius of the
atom ($a = 0.53 \times 10^{-10}~{\rm{m}}/Z^{1/3}$) as
$\chi_a^2={\frac{\lambda}{2\pi a}}^2(1.13+3.76\alpha^2)$ with
$\alpha=\frac{2\pi zZe^2}{hv}$. Then the $B$ parameter of
Eq.~\ref{eq:eq1} is determined by the transcendental equation:
\begin{equation}
B = -\ln\left(\frac{\chi_a^2}{\chi_c^2} \frac{\hat{\gamma}^2}{\hat{e}B}\right)
\end{equation}
where $\hat{e}$ and $\hat{\gamma}$ here are the Euler and Euler-Mascheroni numerical constants.
The angle $\chi$ appearing in Eqns.~\ref{eq:eq1} is $\chi=\frac{\theta}{\chi_c\sqrt{B}}$ .
In the specific case of a 1mm thick Pb foil hit by a 10 GeV muon we find 
$\chi_c= 0.24$~mrad, $\chi_a=1.61~\mu$rad and $B = 14.7$.
The coefficients $c_i$ for a nucleus with nuclear charge density $q(x)$ have the following expressions:
\begin{equation}
c_0 =0.423+2\int_0^{\infty}(1-q(x))(1-e^{-x^2})x^{-3}dx
\end{equation}
\begin{equation}
c_1 =-1.423-2\int_0^{\infty}(1-q(x))e^{-x^2}x^{-1}dx
\end{equation}
\begin{equation}
c_{2\nu} = \frac{2}{(\nu!)^2}\int_0^\infty q(x) x^{2\nu-3}e^{-x^2}dx
\label{eqA}
\end{equation}
If a uniform charged sphere is assumed as in Eq.~\ref{eq:eq3} and 
approximating the resulting form factor with a gaussian function $q=e^{-ax^2}$,
the expression of the coefficients assumes this analytical form: 
\begin{equation}
c_0 =0.423-\left(1-\frac{1}{Z}\right)\left[  \frac{a}{x_n^2}\ln \frac{a}{x_n^2}-\left(1+\frac{a}{x_n^2}\right)\ln \left(1+\frac{a}{x_n^2}\right) \right]
\end{equation}
\begin{equation}
c_1 =-1.423-\left(1-\frac{1}{Z}\right)\ln\left(1+\frac{a}{x_n^2}\right)
\end{equation}
\begin{equation}
\label{eqB}
c_{2\nu} =\frac{(\nu-2)!}{(\nu!)^2}\left[ \frac{1}{Z}+\left(1-\frac{1}{Z}\right) \left(1+\frac{a}{x_n^2}\right)^{1-\nu}\right]
\end{equation}

\end{document}